\documentclass[aps,prl,twocolumn]{revtex4}
\newcommand{\be}{\begin{equation}}
\newcommand{\ee}{\end{equation}}
\newcommand{\bea}{\begin{eqnarray}}
\newcommand{\eea}{\end{eqnarray}}
\usepackage{graphics}
\usepackage{graphicx,epsfig}
\usepackage{amsfonts}
\usepackage{amssymb}
\begin{document}
\title{Gravitational field of a global defect}

\author{D. Bazeia,$^a$ C. Furtado,$^a$ and A.R. Gomes$^{b}$}

\affiliation{ $^a$Departamento de F\'\i sica, Universidade Federal
da Para\'\i ba, Caixa Postal 5008, 58051-970, Jo\~ao Pessoa
Para\'\i ba, Brazil
\\
$^b$Departamento de Ci\^encias Exatas, Centro Federal de
Educa\c c\~ao Tecnol\'ogica do Maranh\~ao, 65025-001 S\~ao Lu\'\i s
Maranh\~ao, Brazil}

\date{\today}
\begin{abstract}
Global topological defects described by real scalar field in (3,1)
dimensions coupled to gravity are analyzed. We consider a class of
scalar potentials with explicit dependence with distance, evading
Derrick's theorem and leading to defects with spherical symmetry.
The analysis shows that the defects have finite energy on flat
space, contrary to the observed for the global monopole. With the
aim to study the gravitational field produced by such defects,
after an {\it Ansatz} for the static metric with spherical
symmetry, we obtain the coupled system of Einstein and field
equations. On the Newtonian approximation, we numerically find
that the defects have a repulsive gravitational field. This field
is like one generated by a negative mass distributed on a
spherical shell. In the weak gravity regime a relation between the
Newtonian potential and one of the metric coefficients is
obtained. The numerical analysis in this regime leads to a
spacetime with a deficit solid angle on the core of the defect.
\end{abstract}
 \maketitle

\section{I.INTRODUCTION}

Topological defects are solutions of the classical field
equations, which are similar to particles.
In quantum theory, they correspond to extended particles, composed
of the elementary particles in each particular model\cite{rb}.
Topological defects are found very frequently in condensed matter
physics. For instance, the vortex\cite{no}, the simplest soliton
in gauge theory with scalars occur in type-2 superconductors. In
the cosmological context we have cosmic strings, and some models
use them as seeds for the formation of structures in the early
universe\cite{vs}. In this way topological defects are of great
interest in high energy physics\cite{vs}.

Topological defects can be global or local. The global defects
arise in models with a continuum but global symmetry, and have
been investigated for instance in\cite{barvil,harlous}. The local
defects usually arise in models which develop spontaneous symmetry
breaking of some local symmetry.

As is well known, however, there is a theorem which puts a limit
on the dimensionality of systems constructed only with scalar
fields, in order to ensure the presence of topological
defects{\citep{der}}. Along the years, several distinct routes
have been constructed to evade this theorem, such as: i) to
consider constraints on the scalar fields, as in the nonlinear
$O(3)$ model\cite{bp}; ii) to add gauge fields, as with the
'tHooft-Polyakov monopole\cite{thooft, poly}; iii) to include
higher-order derivatives on the Lagrangian\cite{afz}; iv) to
introduce time-dependent and non-dissipative solutions (see
ref.\cite{lee} pp.141-150 and ref\cite{lee2}); v) to suppose
non-local solutions, such as global monopoles - this means models
with a divergent energy that can be limited by a cutoff in a
cosmological context due to the presence of another
defect\cite{barvil,harlous}.

Recently, another possibility was considered on searching for
defects described by only one scalar field in $(3,1)$ dimensions.
For this were introduced models described by a dimensionless
Lagrangian density with an explicit dependence with
distance{\cite{bmm}}:
\be
 {\cal{L}}=
\frac12\partial_{\mu}\tilde\phi\partial^{\mu}\tilde\phi -
\frac1{2}f(\tilde x^2)\biggl(\frac{\partial
W}{\partial\tilde\phi}\biggr)^2, \ee where $f(\tilde x^2)$ is in
principle an arbitrary function of the dimensionless coordinate
$\tilde x^\mu$, $\tilde\phi$ a real dimensionless scalar field
  and $W=W(\tilde\phi)$
 a smooth function of $\tilde\phi$.

The equation of motion for static field is, for a specific choice
of $f(\tilde x^2)$,
\be
\label{c3_globr_eqmov}
\partial_i\partial^i\tilde\phi =\nabla^2\tilde\phi =
\frac1{\tilde r^4}\frac{\partial W}{\partial\tilde\phi}
\frac{\partial^2 W}{\partial\tilde\phi^2}, \ee  with $\tilde
r=(\tilde x_1^2+\tilde x_2^2+\tilde x_3^2)^{1/2}$
  a radial dimensionless quantity. Following
Ref.\cite{bmm} we look for solutions depending only on the radial
coordinate $\tilde r$, we change the variable $\tilde r\to 1/x$,
obtaining $\nabla^2\tilde\phi = 1/\tilde r^4 ({d^2 \tilde\phi}/{d
x^2})$. Substituting this in Eq.(\ref{c3_globr_eqmov}), we have
\be
\frac{d^2 \tilde\phi}{d x^2} = \frac{\partial
W}{\partial\tilde\phi} \frac{\partial^2 W}{\partial\tilde\phi^2},
\ee a 1-dim equation on the $x$ variable. Note that the solutions
of the first-order differential equation
\be
\label{c3_globr_bog} 
\frac{d\tilde\phi}{dx} = \frac{\partial W}{\partial\tilde\phi} \ee
are also solutions of the equation of motion.

The class of models
\be
\label{Vppotential}
V_p(\tilde\phi)=\frac12\tilde\phi^2(\tilde\phi^{-\frac1p}-\tilde\phi^{\frac1p})^2
= \frac12\biggl(\frac{\partial W}{\partial\tilde\phi}\biggr)^2\ee
is described by the parameter $p$ which drives the way the field
self-interacts. For $p=1$ we have the usual $\tilde\phi^4$ theory.
For $p=3,5,...$ the models describe potentials supporting minima
at $\tilde\phi=0$ and $\pm1$, and the solutions of
Eq.(\ref{c3_globr_bog}) connect the minima $\tilde\phi=\pm1$
crossing $\tilde\phi=0$, on the form of a 2-kink on the $x$
variable. It was shown in \cite{bmm} that the $p=4,6,...$ cases,
where the potential has minima at $\tilde\phi=0$ and $1$, lead to
stable solutions for Eq.(\ref{c3_globr_bog}), valid for
$\tilde\phi \in [0,\infty)$,  in the form
\be
\label{c3_globr_phir_flat} \tilde\phi(\tilde r) =
\tanh^p\biggl[\frac1{p\tilde r} \biggr]. \ee These solutions with
even $p$, $p\ge 4$, have spherical symmetry, with a central core
growing with $p$. Here we are interested on analyzing the
gravitational properties of one of these solutions with spherical
symmetry. We fix from here on $p=4$, whose solution on the flat
space given by Eq.(\ref{c3_globr_phir_flat}), leads to a smooth
interpolation between the minima $\tilde\phi=1$ and $\tilde\phi=0$
for $\tilde r$ going from $0$ to $\infty$, respectively.
\begin{figure}[ht!]
\includegraphics[{angle=270,width=7.5cm}]{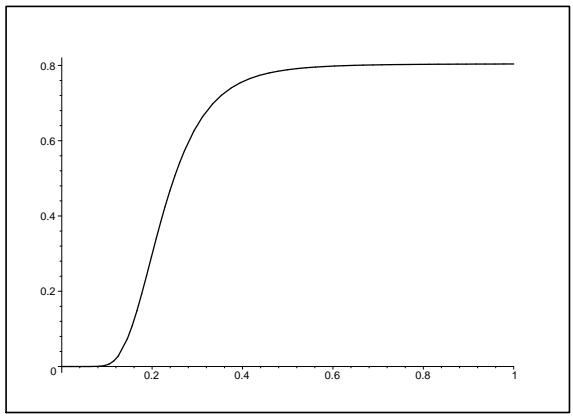}
\caption{Total energy at a distance $\tilde R$ for a defect $p=4$
in flat space, as a function of $\tilde R$.}\label{fig1_new}
\end{figure}

For this case we calculate the energy of the configuration at a
distance $\tilde R$ from the defect. The result, obtained for flat
space and depicted in Fig.\ref{fig1_new}, shows that the energy of
the defect grows until a plateu is achieved. In this way there is
no the characteristic divergence obtained for the global monopole.
This shows that this defect, contrary to the global monopole, does
not need a distance cutoff in order to achieve finite energy.

As this defect was obtained in flat space, a next step is to
analyze its gravitational field. With this aim in Sec. II we
obtain the Einstein and field equations. We show that the system
of equations need four boundary conditions, two of them related to
the behaviour of the scalar potential at the center of the core
and at the infinity and two other ones coming from imposing a
Minkowski spacetime far from the defect. The possible ways to
solve completely this system of equations are discussed. In Sec.
III we consider the Newtonian approximation which shows the
repulsive character of the defect. We also present an analysis of
the weak field approximation, leading as a result that the metric
inside the core of the defect presents a deficit solid angle. Our
conclusions are presented in Sec.IV.

\section{II. EINSTEIN AND FIELD EQUATIONS}

In curved spacetime the action for the defect structure is
\be
S=\int d^4x\,\sqrt{g}\Bigl[-\frac14 R+
\frac12\partial_{\mu}\phi\partial^{\mu}\phi-\frac1{r^4}V(\phi)\Bigr],
\ee with the usual relation between the spherical coordinates
$r,\theta,\phi$ and the Cartesian coordinates $x^i$, $i=1,2,3$ and
where $g=-\det(g_{\mu\nu})$ and $\mu,\nu=0,1,2,3$. The potential
is \be V(\phi) = \frac{1}{\lambda\eta^{5/2}}\frac12
\phi^{3/2}(\eta^{1/2} - \phi^{1/2})^2, \ee which is the potential
from Eq.(\ref{Vppotential}) for $p=4$, after restoring the
dimensional quantity $\eta$, the vacuum expectation value (v.e.v)
of the scalar field $\phi$, and the coupling constant $\lambda$.

As an Ansatz, we try a general static metric with spherical
symmetry:
 \be
\label{ds2D3}
 ds^2=B(r)dt^2-A(r)dr^2-r^2(d\theta^2+sin^2\theta d\phi^2).\ee
 Defining $\tilde\phi = \phi/\eta$ and $\tilde r =
r/\delta$, with $\delta \equiv \lambda^{-1/2}\eta^{-1}$, the
equation of motion for the scalar field can be written as

\begin{eqnarray} \label{c3_globr_eqmov_phi}\nonumber  \frac{d^2\tilde \phi}{d\tilde
r^2} &+& \biggl(\frac1{2B}\frac{dB}{d\tilde r} -
\frac1{2A}\frac{dA}{d\tilde r} + \frac2{\tilde
r}\biggr)\frac{d\tilde \phi}{d\tilde r}\\ &-& \frac A{4\tilde r^4}
(1-\tilde \phi^{1/2})(3\tilde \phi^{1/2} - 5\tilde \phi) = 0
\end{eqnarray}
The
energy-momentum tensor of the defect is related to the variation
of the Lagrangian density of the matter field with respect to the
metric\cite{note1}:
\be
\label{c3_Tmunu} T_{\mu\nu} = \frac 2{\sqrt g}
\frac{\partial\mathcal L_{mat}}{\partial g^{\mu\nu}}.
 \ee
Using the result $\partial g/\partial g^{\mu\nu} = -g g_{\mu\nu}$
(v.\cite{wein}, p.364), we have for a Lagrangian with an usual
$(1/2)\partial_{\mu}\phi^a\partial^{\mu}\phi^a$ kinetic term:
\be
T_{\mu\nu}=\partial_{\mu}\phi^a\partial_{\nu}\phi^a -
g_{\mu\nu}\mathcal L^{flat}_{mat}, \ee where $\mathcal
L^{flat}_{mat}$ is the usual Lagrangian of matter field in flat
space. This gives for our model the following non-null components
for the energy-momentum tensor:
\begin{eqnarray}
T_t^t &=& T_{\theta}^{\theta}=T_{\phi}^{\phi}= \frac{\phi'^2}{2A}
+\frac1{r^4} V(\phi)\\ T_r^r &=& -\frac1{2A} \phi'^2 + \frac1{r^4}
V(\phi),
\end{eqnarray}
where the prime denotes derivation with respect to $r$. The
Einstein equations are \be \label{c3_eqEinstein} R_\mu^\nu-\frac12
R g_\mu^\nu \equiv G_\mu^\nu = 8\pi G T_\mu^\nu, \ee where
$R_\mu^\nu$ and $G_\mu^\nu$ are, respectively, the Riemann and
Einstein tensors.

The components of the Einstein tensor are easily obtained using
the computer algebra systems GRTensorII (for Maple):
\begin{eqnarray}
G_t^t &=& \frac{A'}{A^2r} + \frac1{r^2} - \frac1{Ar^2}, \\
 G_r^r &=& \frac{B'}{ABr} - \frac1{r^2} + \frac1{Ar^2},\\
G_\theta^\theta &=& G_\phi^\phi = \frac{B'}{2ABr} -
\frac{A'}{2A^2r} - \frac{B'A'}{4A^2B}+ \frac{B''}{2AB} \\& & -
\frac{B'^2}{4AB^2}.
\end{eqnarray}

With the components of Einstein and energy-momentum tensors, and
changing to the $\tilde r$ variable we get explicitly:
\begin{eqnarray}
\nonumber \label{c3_globr_eqA} \frac{dA}{d\tilde r} &=& A^2\tilde
r \biggl[\frac1{A\tilde r^2} - \frac1{\tilde r^2} \\ &+& \Delta
\biggl(\frac1{2A}\biggl(\frac{d\tilde\phi}{d\tilde r}\biggr)^2 +
\frac1{2\tilde r^4} \tilde\phi^{3/2} (1-\tilde\phi^{1/2})^2
\biggr) \biggr], \\
 \label{c3_globr_eqB} \frac{dB}{d\tilde r} &=&
-\frac{dA}{d\tilde r}\frac BA + \Delta B \tilde r
\biggl(\frac{d\tilde\phi}{d\tilde r}\biggr)^2 \end{eqnarray} and
\begin{eqnarray}
 \label{c3_globr_eqconsist}  \nonumber\frac{B'}{2ABr}
&-&\frac{A'}{2A^2r} - \frac{B'A'}{4A^2B}+ \frac{B''}{2AB} -
\frac{B'^2}{4AB^2} \\ &=& \Delta
\biggl(\frac1{2A}\biggl(\frac{d\tilde\phi}{d\tilde r}\biggr)^2 +
\frac1{2\tilde r^4} \tilde\phi^{3/2} (1-\tilde\phi^{1/2})^2
\biggr)
 \end{eqnarray}
with $\Delta \equiv {\eta^2}/{M_p^2}$. We see that the Einstein
equations (Eqs.(\ref{c3_globr_eqA})-(\ref{c3_globr_eqconsist}))
together with the equation of motion
(Eq.(\ref{c3_globr_eqmov_phi})) for the $\tilde\phi$ field form a
highly nonlinear set of equations.

The solution $\tilde\phi(\tilde r)$ for flat space
(Eq.(\ref{c3_globr_phir_flat}), with $p=4$) impose the boundary
conditions (b.c.) $\tilde\phi(0)=1$ and $\tilde\phi(\infty)=0$.
Also, as the matter content (related with the $\phi$ field)
decreases with distance, we expect that far from the defect the
spacetime tend to be four-dimensional Minkowski, $M_4$. This leads
to the additional conditions $A(\infty)=1, B(\infty)=1$.

From the b.c. we see that the set of equations for $A$, $B$,
$\tilde\phi$ is over-determined, and one of the Einstein equations
is used as a consistency check. In a numerical search for a
complete set of solutions the b.c. favor  the use of
Eqs.(\ref{c3_globr_eqA})-(\ref{c3_globr_eqB}) together with
Eq.(\ref{c3_globr_eqmov_phi}). The hybrid nature of the b.c.
signals that the use of relaxation (or shooting) method for
Eq.(\ref{c3_globr_eqmov_phi}) together with a Runge-Kutta method
for Eqs.(\ref{c3_globr_eqA})-(\ref{c3_globr_eqB}) can be used in a
search for a solution. This study is in progress. In the present
work, however, as we are interested in a weak field limit, we will
use Eqs.(\ref{c3_globr_eqA}) and (\ref{c3_globr_eqconsist})
together with the solution for the $\phi$ field obtained in flat
space (Eq.(\ref{c3_globr_phir_flat}) for $p=4$).

\subsection{III. LINEARIZED GRAVITY}
\begin{figure}[ht!]
\includegraphics[{angle=270,width=7.5cm}]{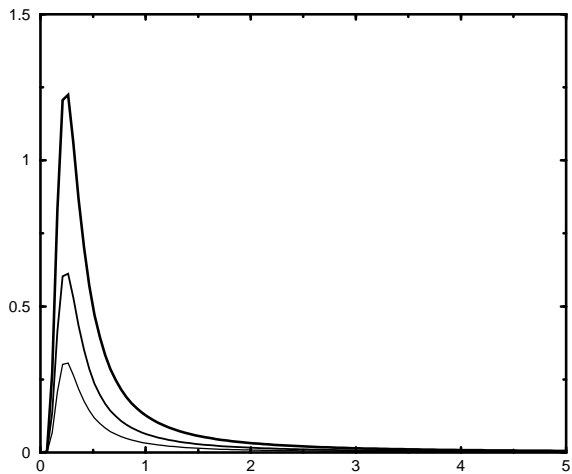}
\caption{Repulsive Newtonian force $F(r/\delta)$ acting on a
         particle with unitary mass for $\Delta= 0.5$,
$1.0$ and $2.0$. Here and in the following figures the thickness
of the lines increases with increasing $\Delta$.} \label{Phir_new}
\end{figure}

Consider first the Newtonian approximation. The Newtonian
gravitational potential $\Phi(r)$,  is obtained as a solution of
the Poisson equation
\be
\label{c3_PhiT0Ti} \nabla^2\Phi = 8\pi G (T_\mu^\mu - \frac12
\eta_\mu^\mu T) = 4\pi G (T_0^0 - T_i^i) . \ee Substituting the
stress-energy tensor on the Poisson equation, after setting $A=1$
for the Newtonian approximation, we get
\be\label{PhiD3} \nabla^2\Phi=-\frac{V(\phi)}{M_p^2r^4} 
 \ee
or, in terms of the dimensionless variables $\tilde\phi$ and
$\tilde r$,
\be
\label{c3_globr_eqPhi} \frac{d^2\Phi}{d\tilde r^2} + \frac2{\tilde
r}\Phi = -\Delta\frac{\tilde\phi^{3/2}}{2\tilde
r^4}(1-\tilde\phi^{1/2})^2. \ee In this approximation (which for
our problem means $\tilde r>>1$), we can consider
$\tilde\phi(\tilde r)$ as the solution on flat space,
$\tilde\phi(\tilde r)=\tanh^4(1/(4\tilde r))$. A graphic of the
r.h.s. of Eq.(\ref{c3_globr_eqPhi}) as a function of $\tilde r$
indicates that the defect acts as a thick spherical shell with
negative mass. In this way, we expect a potential with a constant
value on the core of the defect, with a maximum repulsive force
near to the shell and decreasing with the distance to the shell.
As the potential is defined unless a constant, we numerically
solved Eq.(\ref{c3_globr_eqPhi}) using the second order
Runge-Kutta method, with the initial conditions $\Phi(0)=1$ e
$\Phi'(0)=0$. The corresponding gravitational force under a
particle with unitary mass is depicted in Fig.\ref{Phir_new}. This
figure shows a localized aspect of the gravitational force, with
the magnitude increasing with the parameter $\Delta$.

\begin{figure}[ht!]
\includegraphics[{angle=270,width=7.5cm}]{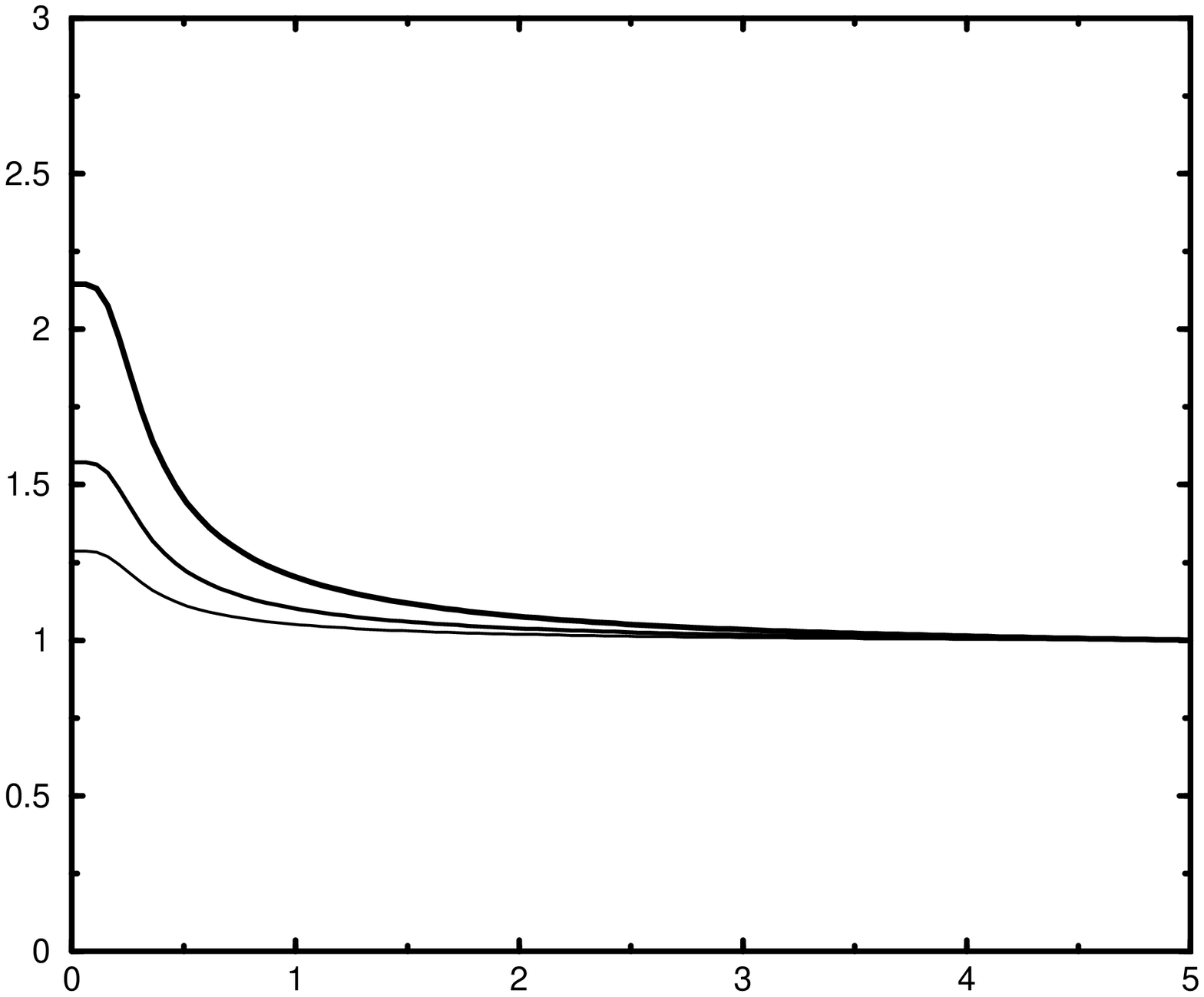}
\includegraphics[{angle=270,width=7.5cm}]{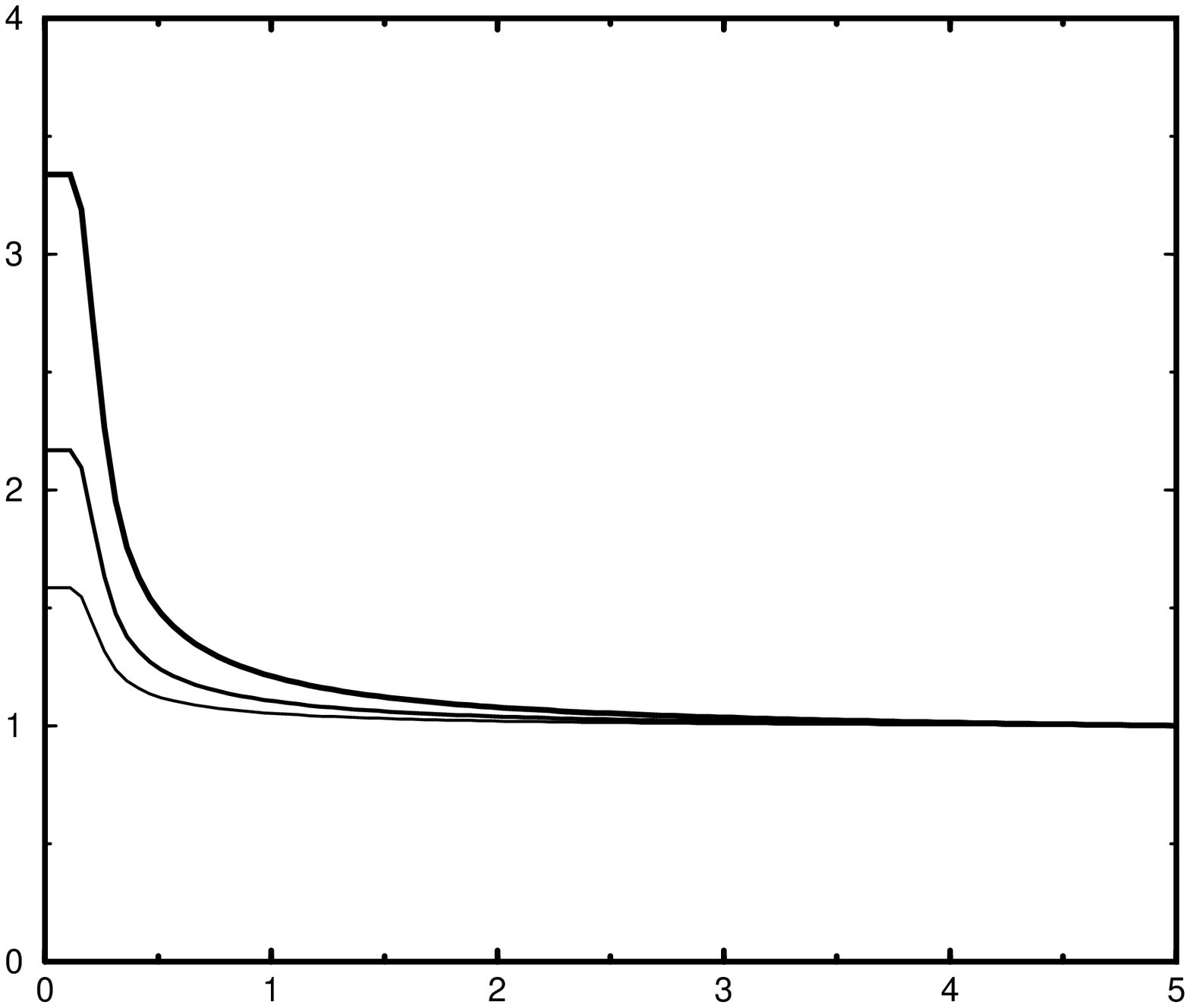}
\caption{Metric coefficients (a)$B(r/\delta)$(up) and
(b)$A(r/\delta)$(down) , on the linear approximation, for $\Delta=
0.5$, $1.0$ and $2.0$.} \label{linAr_new}
\end{figure}

The next step is to investigate the influence of this defect over
the spacetime geometry, on the linear approximation. With this
purpose we consider $A$ and $B$ from Eq.(\ref{ds2D3}) in the form
\be
A(r)=1+\epsilon\alpha(r)\, , \, B(r)=1+\epsilon\beta(r). \ee  The
$\theta\theta$ component of the Einstein equations gives, for this
approximation, \be \frac{d^2\beta}{d\tilde r^2} + \frac2{\tilde r}
\beta= -\Delta\frac{\tilde\phi^{3/2}}{\tilde
r^4}(1-\tilde\phi^{1/2})^2. \ee After comparison with the Poisson
equation for the Newtonian gravitational potential, we are led to
$\beta = 2\Phi + c$, where $c$ is a constant determined with the
boundary condition $B(\infty)=1$. The result is depicted in
Fig.\ref{linAr_new}(a). Also, Eq.(\ref{c3_globr_eqB}) in this
approximation leads to
\be
\frac{d\alpha}{d\tilde r} = - \frac{d\beta}{d\tilde r} +
\Delta\tilde r \bigl( \frac{d\tilde\phi}{d\tilde r}\bigr)^2. \ee
We numerically integrate this equation with $\tilde\phi$ given by
the solution on flat space, and with the result for $\beta(r)$
obtained previously. In this way we obtain Fig.\ref{linAr_new}(b).
Note that Figs.\ref{linAr_new}(a)-(b) show that near to the
origin, the
 solutions for $A$ and $B$ tend to different values, indicating the
 appearance of a deficit solid angle.

\subsection{IV. CONCLUSION}
In this work we analyzed the gravitational field produced by a
global defect with an explicit $1/r^4$ dependence on the scalar
potential. We found that with such a potential the defect produced
has finite energy. This is an alternative to the global
monopole\cite{barvil,harlous} for constructing a $(3,1)$-dim
defect, where no cutoff distance is needed. The analysis of the
Newtonian potential shows no effect besides a repulsive
gravitational field appearing near the core of the defect. The
repulsive Newtonian force is proportional to the v.e.v. $\eta$ of
the scalar field. The analysis of the gravitational field on the
linear regime shows a space with a deficit solid angle. The amount
of this deficit is controlled by the v.e.v. of the scalar field.
In fact, Figs.\ref{linAr_new}(a)-(b) show that inside the defect,
as the parameter $\Delta=\eta/M_p$ increases, the metric
coefficients $A$ and $B$ are more departed from unity. Our weak
field analysis shows the $1/r^4$ dependence of the scalar
potential fixes qualitatively the gravitational effects of the
defect. The magnitude of these effects is governed mainly by the
v.e.v. of the scalar potential. We expect that a further numerical
analysis of the complete set of Einstein and field equations can
reveal other aspects not presented in this work.

The authors would like to thank PADCT/CNPq and PRONEX/CNPq/FAPESQ for financial
support. DB and CF also thank CNPq for partial support.



\begin{thebibliography}{99}
\bibitem{note1}{Note that $\mathcal L_{mat}$ differs from the usual
Lagrangian of matter field in flat space $\mathcal L^{flat}_{mat}$
by the $\sqrt{g}$ term.}
\end{thebibliography}
\end{document}